\documentclass[aps,prl,twocolumn,showpacs,floatfix]{revtex4}
\pdfoutput=1

\usepackage{amsmath}
\usepackage{amssymb}
\usepackage{graphicx}
\usepackage{epstopdf}

\begin{document}

\newcommand{\bmath}{\begin{displaymath}}
\newcommand{\emath}{\end{displaymath}}

\newcommand{\be}{\begin{equation}}
\newcommand{\ee}{\end{equation}}
\newcommand{\bea}{\begin{eqnarray}}
\newcommand{\eea}{\end{eqnarray}}
\newcommand{\non}{\nonumber\\}
\newcommand{\bmultl}{\begin{multline}}
\newcommand{\emultl}{\end{multline}}

\newcommand{\bsubeq}{\begin{subequations}}
\newcommand{\esubeq}{\end{subequations}}
\newcommand{\bitemize}{\begin{itemize}}
\newcommand{\eitemize}{\end{itemize}}
\newcommand{\ket}[1]{\left|{#1}\right\rangle}
\newcommand{\bra}[1]{\left\langle{#1}\right|}
\newcommand{\abs}[1]{\left|{#1}\right|}
\newcommand{\re}{\mathrm{Re}}
\newcommand{\im}{\mathrm{Im}}
\newcommand{\bmx}{\begin{bmatrix}}
\newcommand{\emx}{\end{bmatrix}}
\newcommand{\bsmx}{\begin{smallmatrix}}
\newcommand{\esmx}{\end{smallmatrix}}

\def\ud{\mathrm{d}}
\def\dt{\frac{\partial}{\partial t}}
\def\R{\vec{\hat{R}}}
\renewcommand{\Re}{\mathrm{Re}}
\renewcommand{\vec}[1]{\underline{#1}}
\newcommand{\mat}[1]{\mathbf{#1}}
\newcommand{\rev}[1]{\vec{r}^{(#1)}}
\newcommand{\lev}[1]{\vec{l}^{(#1)}}

\newcommand{\bquote}{\quotedblbase{}}
\newcommand{\equote}{\textquotedblright{ }}

\title{The Dicke model phase transition in the quantum motion of a Bose-Einstein condensate in an optical cavity}
\author{D. Nagy$^{1}$}
\author{G. K\'onya$^{1}$}
\author{G. Szirmai$^{1,2}$}
\author{P. Domokos$^{1}$}

\affiliation{$^{1}$Research Institute for Solid State Physics and Optics, H-1525 Budapest P.O. Box 49, Hungary}
\affiliation{$^{2}$ICFO-Institut de Ci\`encies Fot\`oniques, 08860 Castelldefels (Barcelona), Spain}
%
%
\begin{abstract}
We show that the motion of a laser-driven Bose-Einstein condensate in a high-finesse optical cavity realizes the spin-boson Dicke-model. The quantum phase transition of the Dicke-model from the normal to the superradiant phase corresponds to the self-organization of atoms from the homogeneous into a periodically patterned distribution above a critical driving strength. The fragility of the ground state due to photon measurement induced back action is calculated.
\end{abstract}


\maketitle

A thermal cloud of cold atoms interacting with a single mode of a high-finesse optical cavity can undergo a phase transition when tuning the power of a laser field which illuminates the atoms from a direction perpendicular to the cavity axis \cite{domokos02b,asboth05,nagy06,black03}. Below a threshold power, the thermal fluctuations stabilize the homogeneous distribution of the cloud, and photons scattered by the atoms into the cavity destructively interfere, rendering the mean optical field to be zero. Above threshold, the atoms self-organize into a wavelength-periodic crystalline order bound by the radiation field which, in this case, is composed of the constructive interference of photons scattered off the atoms from the laser into the cavity.  The same phase transition can happen for Bose-Einstein condensed ultra-cold atoms, that is exempt from thermal fluctuations. For low pump power at zero temperature, the homogeneous phase is stabilized by the kinetic energy and the atom-atom collisions,  a sharp transition threshold is thus expected \cite{nagy08,vukics07b}. 
In both examples the self-organization is a non-equilibrium phase transition with the distinct phases being stationary states of the driven-damped dynamics. 

In this paper we show that the Hamiltonian underlying the spatial self-organization is analogous to the Dicke-type Hamiltonian \cite{Dicke1954Coherence} and the transition to the self-organized phase can thus be identified with the superradiant quantum phase transition \cite{Emary2003Chaos}. Hence, the quantum motion of ultracold atoms in a cavity effectively realizes the Dicke model and may lead to the first experimental studies on this paradigmatic system. The accessibility of such a Hamiltonian dynamics is limited by the coupling to the environment. We explore how quantum noise infiltrates and depletes the ground state \cite{murch08}, imposing thereby a condition on the time duration allowed for the adiabatic variation of the macroscopically populated ground state  by means of tuning an external parameter. 

We consider a zero-temperature Bose-Einstein condensate of a number of $N$ atoms of mass $m$ which is inside a high-{\it Q} optical cavity with a single quasi-resonant mode of frequency $\omega_C$. Such a system has been realized and manipulated in several recent experiments \cite{ottl05,Brennecke2008Cavity,Brennecke2008CavityO,slama:063620,klinner06,colombe07}.  The atoms are coherently driven from the side by a pump laser field. The pump laser frequency $\omega$ is detuned far below the atomic resonance frequency $\omega_A$, so that the atom-pump (red) detuning $\Delta_A = \omega - \omega_A$ far exceeds the rate of spontaneous emission.  One can then adiabatically eliminate the excited atomic level and the atom acts merely as a phase-shifter on the field. The dispersive atom-field interaction has a strength $U_0 = g_0^2/\Delta_A$, where $g_0$ is the single-photon Rabi frequency at the antinode of the cavity mode. We describe the condensate dynamics in one dimension along the cavity axis ${x}$, where the cavity mode function is $\cos\,kx$. The motion perpendicular to the cavity axis requires a trivial generalization of the theory, and with a standing-wave side pump the self-organization effect occurs quite similarly in two-, and three dimensions \cite{asboth05}.

The many-particle Hamilton operator in a frame rotating at the pump frequency $\omega$ and with $\hbar =1$ reads
\begin{multline}
\label{eq:H_total}
H = -\Delta_C\,a^\dagger{}a 
 + \int_0^L \Psi^\dagger(x)\bigg[-\frac{\hbar}{2\,m}\frac{d^2}{dx^2}  \\
  + U_0\,a^\dagger{}a\cos^2(kx) + i \eta_t \cos{kx} (a^\dagger -a)\bigg]\Psi(x)dx,
\end{multline}
where $\Psi(x)$ and $a$ are the annihilation operators of the atom field and
the cavity mode, respectively. The cavity length is $L$, the detuning $\Delta_C = \omega -
\omega_C$ is the effective photon energy in the cavity.  Atom-atom s-wave scattering is neglected. Besides the
dispersive interaction term $U_0\cos^2kx$, there is another sinusoidal atom-photon coupling term describing an effective cavity-pump with the amplitude $\eta_t = \Omega{}g_0/\Delta_A$, where $\Omega$ is the Rabi frequency of the coupling to the transverse driving field.

Self-organization is a transition from the homogeneous to a $\lambda$-periodic distribution. The minimum Hilbert-space for the atom field required to describe this transition is spanned by two Fourier-modes,
\be
\label{eq:state_ansatz}
\Psi(x) = \frac{1}{\sqrt{L}} c_0 + \sqrt{\frac{2}{L}} c_1 \cos{k x}\,,
\ee
where $c_0$ and $c_1$ are bosonic annihilation operators. In the low excitation regime these two modes can be assumed to form a closed subspace, so $c_0^\dagger c_0 + c_1^\dagger c_1=N$ is a constant of motion giving the number of particles.
On  invoking the Schwinger-representation in terms of the spin $\hat S$ with components $\hat S_x = \frac{1}{2} (c_1^\dagger c_0 + c_0^\dagger c_1)$,  $\hat S_y = \frac{1}{2i} (c_1^\dagger c_0 - c_0^\dagger c_1) $ and the population difference $\hat S_z = \frac{1}{2} (c_1^\dagger c_1 - c_0^\dagger c_0)$,
the Hamiltonian Eq.~(\ref{eq:H_total}) confined into the two-mode subspace reads
\begin{multline}
H = -\delta_C\,a^\dagger{}a + \omega_R \hat S_z  + i y (a^\dagger -a ) {\hat S_x}/{\sqrt{N}}\\ + u a^\dagger{}a \left( \tfrac{1}{2} + {\hat S_z}/{N} \right) \; , 
\end{multline}
where $\delta_C = \Delta_C - 2 u$, $\omega_R= \hbar k^2/2m$, $u= N\, U_0/4$, and $y= \sqrt{2 N} \eta_t$. In the first line one can recognize the famous Dicke-model Hamiltonian with a coupling constant $y$ tunable via the transverse driving amplitude $\eta_t$. The last term is inherent to the BEC-cavity system, however, it does not essentially change the conclusions to be drawn here as long as $|u|<|\delta_c|$. This condition has to be anyway fulfilled so that the neglect of the next excited Fourier-mode $c_2 \cos{2kx}$ be justified in Eq.~(\ref{eq:state_ansatz}). In the following, we will restrict the discussion to the parameter regime which is needed for the self-organization \cite{nagy08}. That is,  $\delta_C < 0$  is required in order to avoid a dynamical instability of the system, which arises from the motional heating induced by the slightly retarded cavity field \cite{domokos03}. 

The thermodynamic limit is defined as $N \rightarrow \infty$ and $V \rightarrow \infty$, while the atom density  $\rho \propto N/V$ is kept constant. The coupling constants $u$ and  $y$ have been introduced such that they are proportional to the atom density, $u \propto N/V$ and $y\propto \sqrt{N/V}$ (there is a filling factor coefficient), and thus remain constant  in the thermodynamic limit.  The ground state can be determined as in Ref.~\cite{Emary2003Chaos}. Let's use the Holstein-Primakoff representation in which the spin-$N/2$ degree of freedom is expressed in terms of the bosonic operator $b$ such that $\hat S_-=\sqrt{N- b^\dagger b}\, b$, $\hat S_+=b^\dagger\, \sqrt{N- b^\dagger b}$, and $\hat S_z= b^\dagger b- N/2$. The Hamiltonian transforms into
\begin{multline}
\label{eq:H_HP}
H = -\delta_C\,a^\dagger{}a + \omega_R b^\dagger{}b  + u a^\dagger{}a b^\dagger b/N \\
+ \frac{i}{2} y (a^\dagger -a ) \left(b^\dagger\, \sqrt{1- \frac{b^\dagger b}{N}} + \sqrt{1- \frac{b^\dagger b}{N}}\, b\right)  \; .
\end{multline}
Next, let's employ the similarity transformation $\hat D^{-1}(\beta) \hat D^{-1}(\alpha) {H} \hat D(\alpha) \hat D(\beta)$, with the displacement operators, $\hat D(\alpha)=\exp\{ \alpha a^\dagger - \alpha^* a\}$ and $\hat D(\beta)=\exp\{ \beta b^\dagger - \beta^* b\}$, which does not change the spectrum of the Hamiltonian. Formally, the transformation amounts to replacing $ b \rightarrow b + \beta$, $a \rightarrow a + \alpha$, and analogously for the hermitian adjoint operators in (\ref{eq:H_HP}). The resulting Hamiltonian is then expanded up to second-order in the boson operators. Note that the expansion of the nonlinear square root term can be performed only approximately, because the physically sensible Hilbert space for the operator  $b$ is truncated as $b^\dagger b < N$. Therefore the forthcoming results are exact up to $1/N$. 

\begin{figure}
\centering
\includegraphics[angle=0,width=0.85\columnwidth]{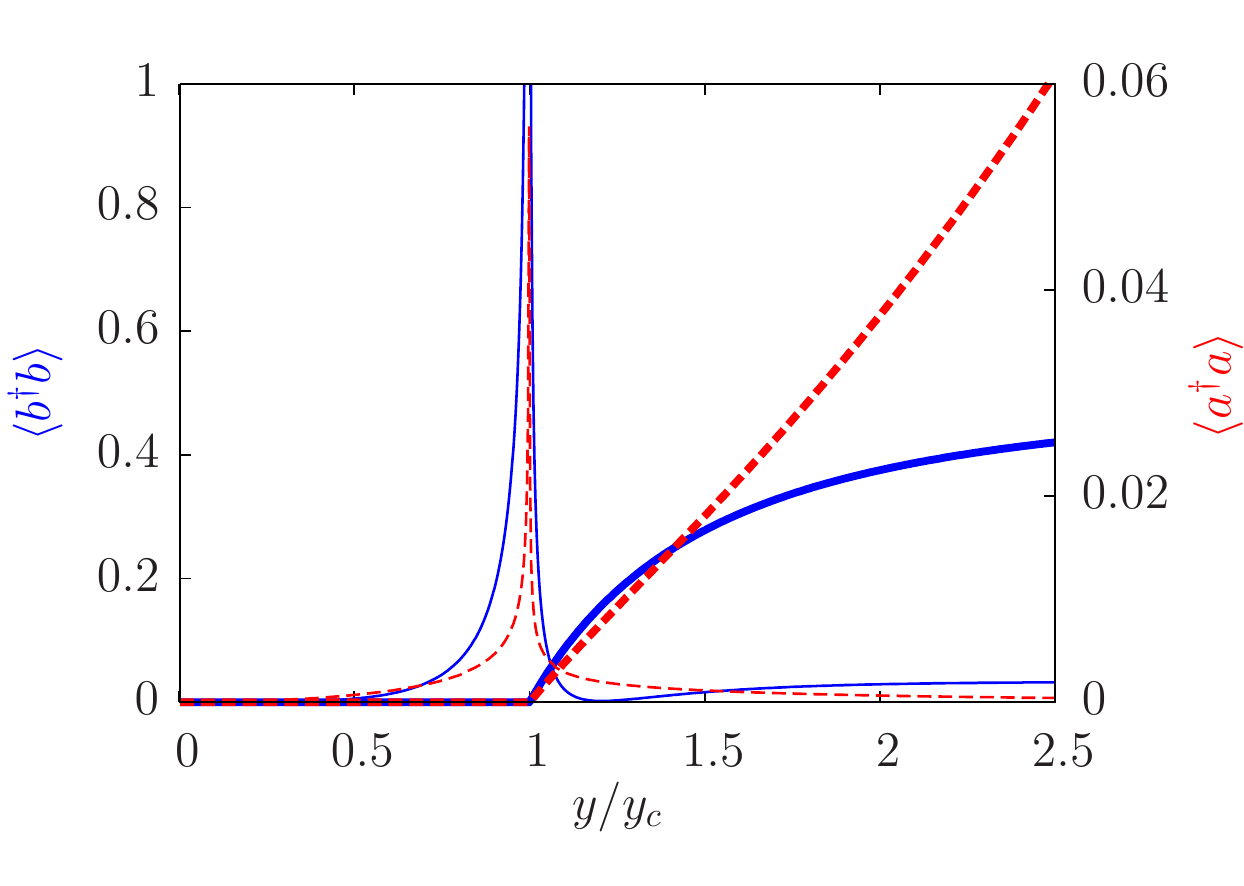}
\caption{(Color online) Photon (dashed red lines) and motionally excited atom (solid blue lines) numbers in the ground state.  Thick lines represent the contributions form the mean fields ($\alpha_0^2$ and $\beta_0^2$, these are the photon and atom excitation numbers divided by $N$, respectively), which can be the order parameters of the phase transition. Thin lines represent the incoherent excitations due to the squeezing, given by the quantum averages  $\langle a^\dagger a \rangle$ and $\langle b^\dagger b\rangle$ taken in the ground state.  Parameters: $\delta_C=-100 \omega_R$, $u=-0.1\omega_R$.  }
\label{fig:qstat_mean}
\end{figure}

There is a pair of real $\beta_0$ and $\alpha_0$ such that the linear terms in the Hamiltonian in the displaced phase space vanish for $\alpha=i\sqrt{N}\,\alpha_0$ and $\beta=\sqrt{N}\,\beta_0$. They obey
\begin{subequations}
\label{eq:cond}
\begin{align}
\left( {\delta}_{C} - u {\beta_0^2} \right) {\alpha_0}
&= y \: \beta_0 \: \sqrt{1 -{\beta_0^2}} \,, \\
\left( \omega_{R} +u {\alpha_0^2}  \right) \beta_0
&= - y \: {\alpha_0} \:
\frac{1  - 2 \beta_0^2 }{\sqrt{1  -\beta_0^2}}\,.
\end{align}
\end{subequations}
The trivial solution $\alpha_0=\beta_0=0$ always satisfies these equations, which corresponds to the physical state of a homogeneous condensate and no photon in the cavity.  When $\alpha_0\neq 0$ and $\beta_0 \neq 0$,  
the product of the two equations leads to a second-order algebraic equation,
\begin{equation}
\label{eq:2ndorder}
\frac{u}{\delta_C} \: \beta_0^4 - 2 \: \beta_0^2 +  \frac{\delta_C\omega_R+y^2}{u\omega_R+y^2} =0 \;�,
\end{equation}   
where we used that $\delta_C \, ( u\omega_R+y^2)\neq 0$. 
There is a physically sensible solution in the range $0< \beta_0^2\leq1$ if and only if $y > y_{\rm crit} \equiv  \sqrt{-\delta_C \omega_R}$.
Then, 
\be
\label{eq:x}
 \beta_0^2 = \frac{\delta_C}{u} \left(1-\sqrt{1- \frac{u}{\delta_C}  \frac{y^2-y_{\rm crit}^2}{y^2 - \tfrac{u}{\delta_C} y_{\rm crit}^2}} \right) \; .
\ee
For $u=0$, which amounts to the normal Dicke model, the solution is $\beta_0^2=\frac{y^2-y_{\rm crit}^2}{2y^2}$ with the same critical value $y_{\rm crit}$ of pump amplitude. The light shift term does not influence the threshold, because the zero mean fields make this term vanish below threshold. Note also that this result for $y_{\rm crit}$ corresponds to the one calculated from the instability of the Gross-Pitaevski equation (GPE) \cite{nagy08}, if this latter is taken in the $g_c \rightarrow 0$ (collisionless atoms) and $\kappa \rightarrow 0$ (no cavity loss) limit. The approach based on the GPE was exempt from the two-mode approximation. On the other hand, owing to the simplicity of the two-mode model, the exact ground state, which might include macroscopic quantum correlations, can be well approximated from the Hamiltonian obtained up to the quadratic order:
\begin{multline}
\label{eq:H_2ho}
H=E_0 + M_0 a^\dagger a + \tfrac{M_x + M_y}{2} b^\dagger b \\ +  \tfrac{M_x -M_y}{4} \left({b^\dagger}^2 + b^2\right) +  \tfrac{i}{2} M_c (a^\dagger -a) (b^\dagger + b)\;�,  
\end{multline}
\bsubeq
\label{eq:couplings}
\begin{align}
\mbox{where}\quad M_0 & = - \delta_C + u \beta_0^2\; ,\\
 M_x &= \omega_R + u\:\alpha _0^2- y \alpha_0 \beta_0 \frac{3  - 2 \beta_0^2 }{\left(1  - \beta_0^2  \right)^{3/2} }\; , \\
 M_y &= \omega_R + u\:\alpha _0^2- y \alpha_0 \beta_0 \frac{1 }{\left(1 - \beta_0^2 \right)^{1/2} }\; , \\
M_c &= 2 u \alpha_0\beta_0 + y   \frac{1  - 2 \beta_0^2 }{\left(1 - \beta_0^2 \right)^{1/2} } \; .
\end{align}
\esubeq
The ground state is the vacuum state of the normal mode oscillators which have the eigenfrequencies 
\be
 \label{eq:eigenfrequencies}
 \omega_{\pm}^2 = \tfrac{M_0^2+ M_x M_y}{2}  \pm \sqrt{ \tfrac{(M_0^2 - M_x M_y )^2}{4}  + M_0 M_y M_c^2} \;.
\ee
Below threshold, the energy gap to the first excited state vanishes as $\sqrt{1 - y/y_{\rm crit}}$ on approaching the critical point (the exponent is thus 1/2). 
The ground state contains excited Fock states of the uncoupled photon $a$ and atomic $b$ modes  because of the squeezing (with the coefficient $M_x -M_y$) and two-mode squeezing (with the coefficient $M_c$) terms. Below threshold $M_x -M_y = 0$, thus the ground state is simply the two-mode squeezed vacuum of the $a$ and $b$ modes, which is an entangled state \cite{Lambert2004Entanglement,Buzek2005Instability}. The quantum average of the photon and atom motion excitations in the ground state  can be seen in Fig.~\ref{fig:qstat_mean} together with the mean field populations. The squeezing leads to a singular state at the critical point which amounts to a divergence of the incoherent excitations  $\langle a^\dagger a \rangle$ and $\langle b^\dagger b \rangle$. Therefore, the detection of the continuous transition of the mean field amplitudes $\alpha_0$ and $\beta_0$ requires, e.g., homodyne light measurement or the observation of the interference between the homogeneous and the sinusoidal components of the atom field.


The quantum phase transition associated with the ground state of the Dicke-type Hamiltonian must be influenced by the cavity loss. The coupling to the environment amounts to a quantum measurement of the coupled BEC-cavity system \cite{murch08,nagy09}, and has a back action on its state. Therefore, even at zero temperature, the ground state is being depleted, which process can be modeled as a diffusion. We calculate the rate of diffusion out of the ground state in the following. For a compact notation the variables are arranged in a vector $\R\equiv[\hat{a},\hat{a}^\dagger,\hat{b},\hat{b}^\dagger]$. The Heisenberg equations of motion originating from the quadratic Hamiltonian (\ref{eq:H_2ho}) are linear and are driven by quantum noise terms associated with the photon field decay, 
$\tfrac{\partial}{\partial t}\R=\mat{M}\R+\vec{\hat{\xi}}\,,$where the matrix $\mat{M}$ contains the coupling between the bosonic creation and annihilation operators, and the noise source is $\vec{\hat{\xi}}=[\hat{\xi},\hat{\xi}^\dagger,0,0]$. 
The only non-vanishing noise correlation function is $\langle \xi(t) \xi^\dagger(t')\rangle = 2 \kappa \delta(t-t')$, where $2\kappa$ is the photon loss rate. We neglect the dissipative $-\kappa a$ and $-\kappa a^\dagger$ terms,  because we are interested in the \emph{transient} dynamics and not in the stationary regime of the system. Initially, the dominant effect in irreversibly escaping from the ground state can be attributed to the infiltration of quantum noise (a diffusion process).

The left and right eigenvectors $\lev{k}$ and $\rev{k}$, respectively, of $\mat{M}$  can be used to expand the fluctuation vector $\R$ in terms of normal modes:   $\R\equiv\sum_k\hat{\rho}_k\rev{k}$. By use of the orthogonality of the left and right eigenvectors, $(\lev{k},\rev{l})=\delta_{k,l}$, where $(\vec{a},\vec{b})$ is the scalar product, the normal mode amplitudes are obtained as  $\hat{\rho}_k=(\lev{k},\R)$. 
They evolve independently as
\begin{equation} 
\label{tdflu}
\hat{\rho}_k(t)=e^{-i\omega_k t}\hat{\rho}_k(0)+\int_0^te^{-i\omega_k(t-t')}\hat{Q}_k(t') \ud t'\;�,
\end{equation}
where the projected noise is $\hat{Q}_k\equiv(\lev{k},\vec{\hat{\xi}})$.  In the present Hamiltonian problem, the normal modes form hermitian adjoint pairs $\rho_+, \rho_+^\dagger$ with eigenfrequencies $\pm \omega_+$, and $\rho_- , \rho_-^\dagger$ with frequencies $\pm \omega_-$ from Eq.~(\ref{eq:eigenfrequencies}), respectively, where each pair corresponds to one of the normal mode oscillator of the quadratic Hamiltonian in (\ref{eq:H_2ho}). Second order correlations evolve as
\begin{multline}
\label{eq:rhoav}
\left\langle\hat{\rho}_k(t)\hat{\rho}_l(t)\right\rangle=\left\langle\hat{\rho}_k(0)\hat{\rho}_l(0)\right\rangle e^{-i(\omega_k+\omega_l)t}\\
+2\kappa\frac{1-e^{-i(\omega_k+\omega_l)t}}{i(\omega_k+\omega_l)}{l^{(k)}_1}^*{l^{(l)}_2}^*.
\end{multline}
The first term represents the initial condition. The diffusion is due to the second term in which the linear time dependence can be written as being proportional to the $\sin(x)/x$ function, where $x = (\omega_k+\omega_l)\delta t/2$. 

\begin{figure}
\centering
\includegraphics[angle=0,width=0.85\columnwidth]{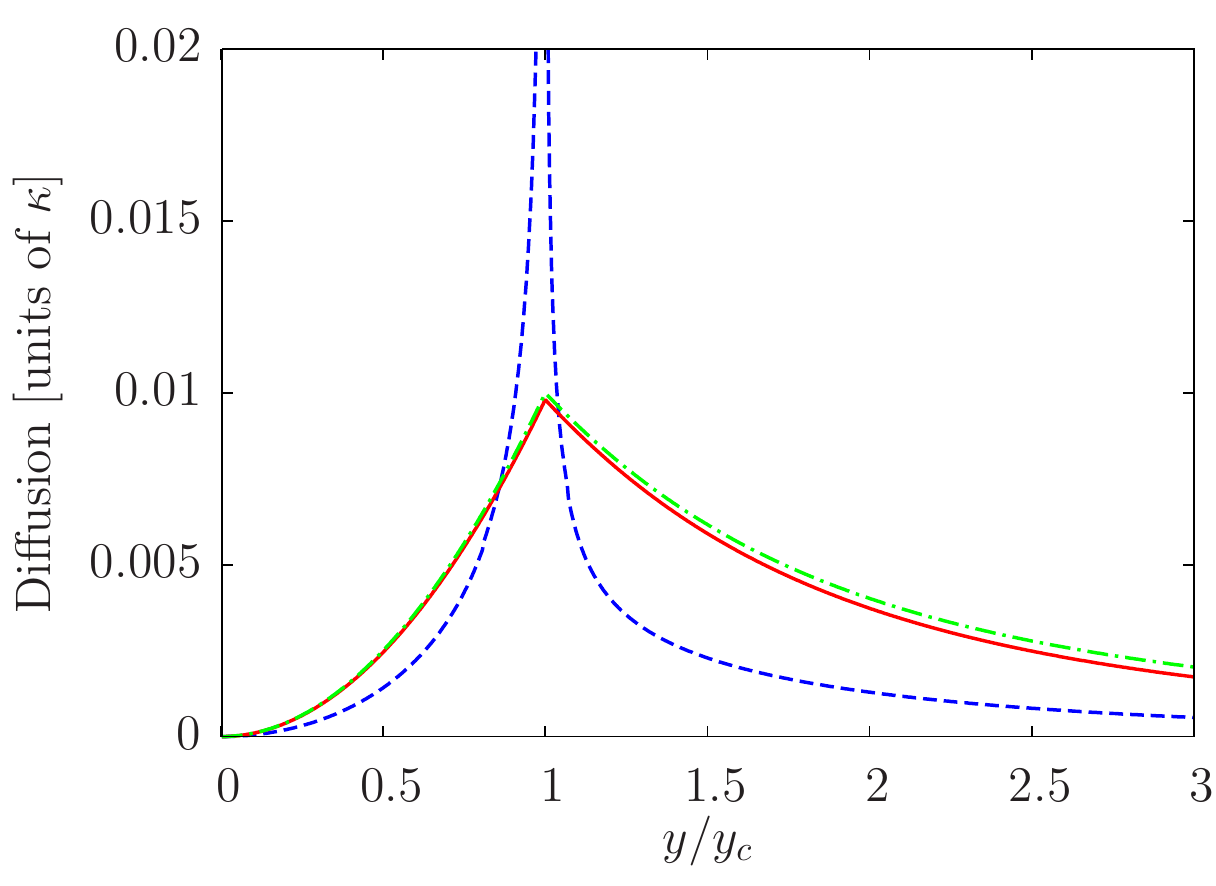}
\caption{Diffusion out from the ground state. The rate of increase of normal mode excitations (dashed line) and that of photons and motionally excited atoms (solid line) with coarse graining  $|\delta_C|^{-1} \ll \delta t \ll \omega_R^{-1}$. The almost overlapping dashed-dotted line is derived from an adiabatic elimination method and is given by the analytic result of Eq.~(\ref{eq:adiabatic_diffusion}). Parameters: $\delta_C=-100 \omega_R$, $u=-0.1\omega_R$. }
\label{fig:diffusion}
\end{figure}

The total population in the excited states above the ground state is given by $\langle \rho_+^\dagger \rho_+ + \rho_-^\dagger \rho_-\rangle$. 
Using (\ref{eq:rhoav}) for its time evolution,  the first term vanishes in the ground state, and, in the second term, $\omega_k+\omega_l=0$ for both $\rho_+^\dagger \rho_+$ and $\rho_-^\dagger \rho_-$ terms. Thus the time evolution leads exactly to a linear increase of the excited population, the corresponding diffusion rate is plotted in Fig.~\ref{fig:diffusion} with dashed line. The singularity at the critical point reflects that the excitation energy of one of the normal modes tends to zero.

Let us calculate the diffusion in terms of measurable quantities, such as the number of incoherent photons and motionally excited atoms, $\delta N = \left\langle a^\dagger a + b^\dagger b \right\rangle$. The incoherent population evolves as
\begin{equation} 
\delta N(t)=\sum_{k,l}\langle\hat{\rho}_k(t)\hat{\rho}_l(t)\rangle \left( r^{(k)}_2 r^{(l)}_1 + r^{(k)}_4 r^{(l)}_3 \right)\; .
\end{equation}
By using Eq.~(\ref{eq:rhoav}) and by approximating the $\sin(x)/x$ function,  the diffusion rate becomes
\begin{multline}
\label{eq:depfin}
\frac{\delta N(t)}{\delta t} \approx 2\kappa\sum_{k,l}
{l^{(k)}_1}^*{l^{(l)}_2}^* \left( r^{(k)}_2 r^{(l)}_1 + r^{(k)}_4 r^{(l)}_3 \right)\\ \Theta\left(\delta t^{-1} - |\omega_k +\omega_l | \right) \; ,
\end{multline}
where $\Theta$ is the Heavyside-function.  If the ``time step'' $\delta t$ is shorter than any of the time periods $\omega_\pm^{-1}$, none of the pairs $(k,l)$ is cut off by the Heavyside-function in the sum~(\ref{eq:depfin}). Then, it follows from the completeness relation $\sum_k r^{(k)}_i {l^{(k)}_j}^* =\delta_{ij}$ that the depletion rate is zero. On such a short time the quantum noise is associated with the photon field amplitudes $a$ and $a^\dagger$, and normal-order products vanish at zero temperature. Diffusion in the populations is obtained when 
a coarse graining of the dynamics over a longer $\delta t$ is performed. We consider only the special case $|\delta_C| \gg \omega_R$, when there is a large difference between the eigenfrequencies $\omega_{\pm}$. The time step can be set such that $|\delta_C|^{-1} \ll \delta t \ll \omega_R^{-1}$, and the two pairs $(k,l)$ with $\omega_k = \omega_l =\pm \omega_-\sim \pm\omega_R$ also contribute to the double sum in  Eq.~(\ref{eq:depfin}), in addition to the $(k,l)$ pairs with $\omega_l=-\omega_k$. Then, the departure from the ground state appears as a regular diffusion process in the motional excitation Fock space with a finite rate even at the critical point, which is plotted in  Fig.~\ref{fig:diffusion} with solid line.

As $\delta_C$ is the far highest frequency in this example, there is an alternative avenue to the depletion rate which relies on the adiabatic elimination of the photon field variables, $a$ and $a^\dagger$, following the method of Ref.~\cite{nagy09}. It leads to an analytical approximation, which is plotted in dashed-dotted line in Fig.~\ref{fig:diffusion},
\begin{equation}
\label{eq:adiabatic_diffusion}
\frac{\delta N(t)}{\delta t}=\kappa \frac{M_c^2}{\delta_C^2+\kappa^2} \;�.
\end{equation}
Below threshold, the diffusion rate is about $\omega_R\, (\kappa/|\delta_C|)\, (y/y_{\rm crit})^2$.  Adiabatic following of the ground state by means of slow variation of the detuning (or the pump amplitude) requires that the smaller excitation frequency, $\omega_- \approx \omega_R \sqrt{1-(y/y_{\rm crit})^2}$, be much larger than the diffusion. The use of large detuning $|\delta_C| \gg \kappa$ {\it removes} the time limitation imposed by the quantum noise. Although the critical point can be adiabatically approached only as close as $y/y_{\rm crit} \ll 1$, which is a generic feature of criticality, it is an intriguing possibility that matter wave and light field entanglement can be adiabatically created by making use of this critical system.

In conclusion, we have shown that the zero temperature limit of the atomic self-organization in a cavity corresponds to the quantum phase transition given by the Dicke model. This connection is principally different from the proposals where some internal electronic dynamics of the atoms in the cloud is involved \cite{Dimer2007Proposed,Chen2008Exotic,larson09,meiser10}. The key point here is that the energies of the decoupled systems are much lower, the atom field excitation is being in the recoil frequency range of kHz, that of the photon field is broadly tunable, and the critical regime can be addressed in currently running experiments \cite{baumann10}. 

We acknowledge funding from the NSF NF68736 and from the NORT (ERC\_HU\_09 OPTOMECH) of Hungary. G.~Sz. acknowledges Spanish MEC projects TOQATA (FIS2008-00784), QOIT (Consolider Ingenio 2010), ERC Advanced Grant QUAGATUA and EU STREP NAMEQUAM.


\end{document}